\documentclass[reprint,amsmath,amssymb,aps]{revtex4-2}

\usepackage{graphicx}
\usepackage{dcolumn}
\usepackage{bm}
\usepackage{float}
\preprint{APS/123-QED}
\begin{document}

\title{Topological State Reconstruction For Wireless Stabilization of Distant Atomic Clocks}

\author{Adam Walton and Benjamin T. H. Varcoe}
 \email{phyawa@leeds.ac.uk}
\affiliation{University of Leeds}

\date{\today}

\begin{abstract}

High-precision frequency alignment with classical communication channels is difficult due to noise, propagation delays, and signal degradation. Current optical methods \cite{1020Clock}, commonly involving frequency combs, are capable of synchronising clocks with exceptional precision up to the region of a part in $10^{20}$. Alternatively, wireless methods see use where this is not practical, with achievable precision within the nanosecond region \cite{GPS}. This leaves few options for achieving high-precision clock synchronisation without requiring specialised equipment, a fibre connection, or a line of sight communication channel. Here we present a novel approach combining quantum state reconstruction with feedback controls to stabilize the frequency of two atomic clocks separated by a 900 MHz free space radio link. Quantum state reconstruction enables tracking of phase and frequency fluctuations during transmission. We see that a part in $10^{16}$ precision in frequency alignment of the clocks can be achieved using commonly-available radio equipment, allowing precise timekeeping and synchronization over long distances provided a radio communications channel can be established, with potential applications in a wide variety of timekeeping applications.
\end{abstract}

\maketitle
Accurate time measurement and synchronization are essential in various scientific and technological applications \cite{Limit,Review}. 
Hence, atomic clocks have proven to be extremely valuable in maintaining precise timekeeping across a wide range of applications, from fundamental physics to navigation. 
Synchronizing atomic clocks has therefore also become a crucial problem in various physics-related fields, such as geodesy \cite{McGrew2018,Mehlstaubler2018,Riehle2017} and fundamental physics measurements \cite{QSNET,Waves,Dark}. Additionally, it plays a vital role in national timekeeping and providing accurate timing information for GPS. In the UK, the National Physical Laboratory (NPL) offers a service with clocks capable of detecting frequency differences on the order of one part in $10^{15}$ over a day \cite{NPL}. 

Precise frequency distribution can be achieved through optical methods provided that placement is practical for fibre cables \cite{Cizek:22} or line-of-sight \cite{Geo,Distant} is available, but for longer distances, wireless methods like Two-Way Satellite Time and Frequency Transfer (TWSTFT) \cite{Space, SynchARS} and Global Navigation Satellite System (GNSS) \cite{GNSS} are employed. However, these wireless methods are less precise, providing time measurements accurate to the nanosecond \cite{NPL}.

In this paper we introduce an innovative approach that combines tomographic state reconstruction with a phase-locked loop (PLL) to stabilize the frequency of two atomic clocks separated by a 900 MHz free-space radio link. Tomographic reconstruction allows the retrieval of information about the phase and amplitude of electromagnetic states, even in the presence of inherent noise and measurement limitations. 
The aim is to detect phase errors in the broadcast channel between two parties using imperfect atomic clocks as frequency references. These phase errors stem from frequency differences between the clocks. By employing these errors as feedback for a controller, we can stabilize the clocks and achieve clock synchronization with impressive precision, achieving drifts below the nano-Hertz level.
Such corrections already happen in ordinary communications systems and form a normal part of device operation, though these corrections are a part of signal processing and do not make changes to the clock source. 
The novel feature introduced here is to correct the frequency of an atomic clock, effectively treating the clock as a flywheel oscillator and thereby allowing the frequency reference to be used by other elements of a system. In particular this would allow for the creation of a wireless "White rabbit" type protocol,

The conventional synchronization of atomic clocks using classical communication channels encounters issues related to noise, propagation delays, and signal degradation. In contrast, combining a topological state reconstruction with a PLL presents a promising solution to overcome these challenges. The PLL is a widely used technique for generating a stable output frequency by comparing it with a reference signal. By incorporating quantum state reconstruction into the PLL, we harness the advantages of quantum measurement techniques to enhance stability and accuracy during the frequency-locking process. This integration opens up new possibilities for achieving precise synchronization between atomic clocks over long distances with unprecedented accuracy.

Topological state reconstruction combined with PLL stabilization of atomic clocks through a free space radio link operating at higher frequencies, such as 2.4 GHz, holds tremendous potential for time synchronization across a network. This approach allows compensation for phase and frequency fluctuations induced by the transmission medium and environmental factors, resulting in improved accuracy and stability. Highly precise synchronization over long distances is crucial for various applications, including global positioning systems, telecommunications, and fundamental scientific research. This paper offers a comprehensive analysis of the proposed system, highlighting its advantages, potential challenges, and experimental results, all of which demonstrate the feasibility and practicality of this approach for future implementations.

\begin{figure}[H]
	\centering
	\includegraphics[width=0.4\textwidth]{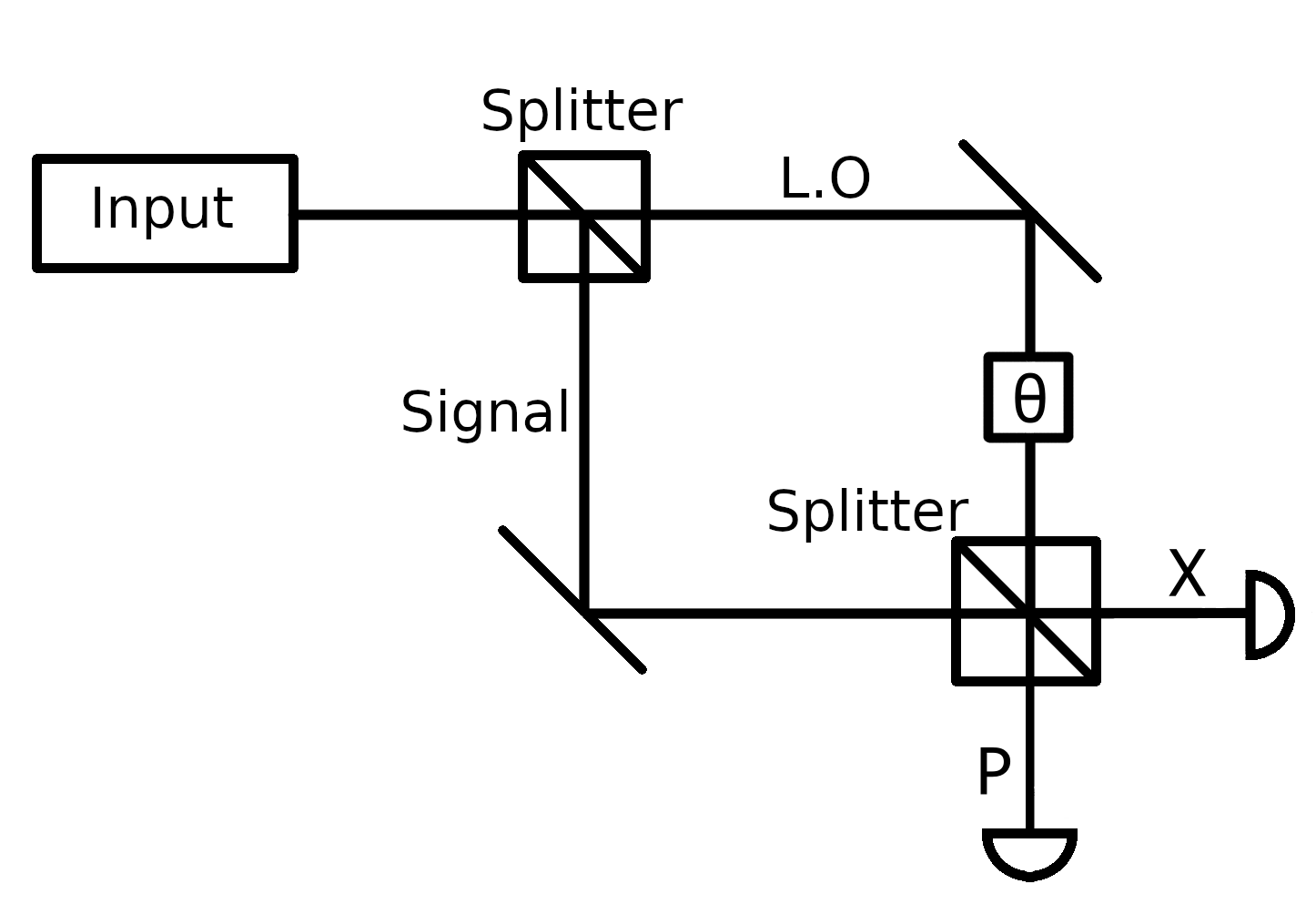}
	\includegraphics[width=0.4\textwidth]{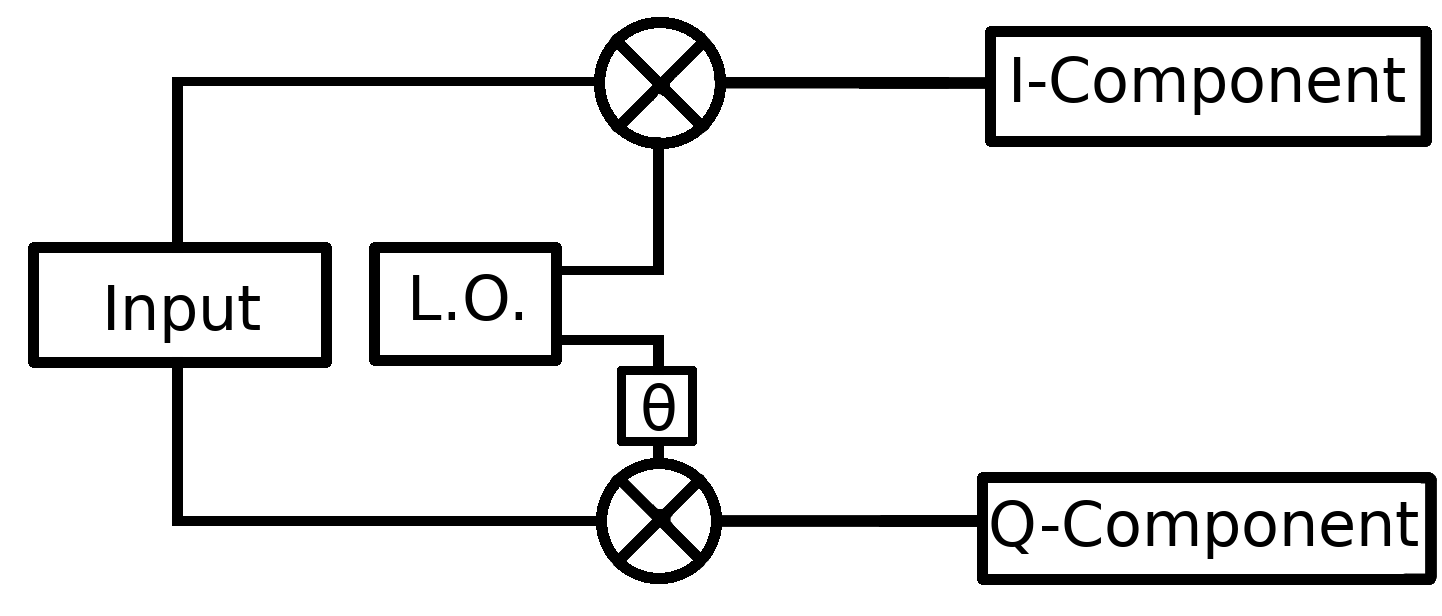}
	\caption{A comparison of an interferometer used for topological state reconstruction (top) to IQ demodulation used for radio signals (bottom). At $\phi=\frac{\pi}{2}$, topological state reconstruction is equivalent to IQ demodulation in radio broadcasting. \label{fig:IQ}}
\end{figure}

We present a measurement of the phase of a 900 MHz microwave link using a homodyne POVM measurement that simultaneously extracts the phase and amplitude of a microwave channel. We use this to align the phase of separated atomic clocks. 
We demonstrate this technique over a free space broadcast channel, achieving a fractional stability of $10^{-16}$ range.  

In Homodyne detection, the signal is split into two parts. Each component is separately detected in a balanced detector measuring the input electric fields $E_s=A_s(t)\exp(i \omega_s t)$ and $E_l=A_l\exp(i \omega_l t)$ via the combinations
\[E_1=\frac{1}{2}(E_s+E_l),~
E_2=\frac{1}{2}(E_s-E_l)\]
\[E_3=\frac{1}{2}(E_s+i E_l),~
E_4=\frac{1}{2}(E_s-i E_l)\]
The $n$th receiver output intensity is therefore given by 
\begin{align*}
    I_n(t) = &\\
R~Re&\left\{\frac{A_s(t)exp(i\omega_s t)+(\pm 1)^n A_l(t)exp(i\omega_l t)}{\sqrt{2}}\right\}
\end{align*}

\begin{align*}
I_1(t) &= 
R[P_s(t)+P_l\\
&+ 2\sqrt{P_s(t)P_l}\cos(\omega_{IF}+\theta_s(t)-\theta_l)]
\end{align*}
\begin{align*}
I_2(t) &= 
R[P_s(t)+P_l\\
&- 2\sqrt{P_s(t)P_l}\cos(\omega_{IF}+\theta_s(t)-\theta_l)]
\end{align*}
where $\omega_{IF}=\omega_s-\omega_l$, $R$ is a normalisation constant, $P_s$ and $P_l$ are respectively the power of the signal and local oscillator and $\theta_s$ and $\theta_l$ are the phase of the signal and local oscillator. 
Typically we make the assumptions that the local oscillator phase and power are both stable and unchanging, removing time dependence. 

The output of the balanced detector is:
\[I_I(t)=I_1(t)-I_2(t)\]
\[\therefore I_I(t) =
R\sqrt{P_s(t)P_l}\cos(\omega_{IF}+\theta_s(t)-\theta_l)]\]
Inserting a $\pi/2$ phase shift into the local oscillator delivers the output of the other quadrature (Fields 3 and 4): 
\[I_Q(t)=I_3(t)-I_4(t)\]
\[\therefore I_Q(t) =
R\sqrt{P_s(t)P_l}\sin(\omega_{IF}+\theta_s(t)-\theta_l)]\]
These two intensities, $I_I(t)$ and $I_Q(t)$, can be viewed as coordinates on the unit circle, allowing us to uniquely determine the phase of the local oscillator.

In two-bit digital transmission, the phase $\theta_s$ is restricted to one of two values $\theta_s\in{0,\pi}$. This is known as binary phase-shift keying (BPSK), the pattern of shifts is called a `cluster'. The benefit of this keying method is that we are able to accurately determine the location of the origin as the center of the cluster and thereby accurately determine the relative phase between the sender and receiver. 
Under this circumstance, the frequency $\omega_{IF}$ will be observed as a rotation of the cluster via the relationship, \[\frac{d\theta_s}{dt}=\omega_{IF}.\]

The experimental method that we have used to synchronise the atomic clocks is to draw a feedback signal from the cluster rotation. The phase-locked loop detects this rotation and feeds this information back to the atomic clock, adjusting the frequency (and phase) to match that of the transmitter. 

Our approach draws parallels with PLLs that are already widely used in digital broadcast channels to ensure consistent frequencies between transmitters and receivers, facilitating smooth radio broadcasting operations. 
However, a crucial distinction lies in two areas: firstly, our use of frequency scaling, where, rather than relying on the 10 MHz reference provided by an atomic clock, we scale up the frequency to 900MHz.
Secondly the aim of the PLL is to stabilise an external clock. This allows us to extend the reach of a single master oscillator to a network of local oscillators. 

The rationale behind this frequency scaling stems from the assumption that any error in the 10 MHz reference is a result of changes in the time measurement of the clocks. As the frequency increases, the phase shift for a given time error also increases. As we are sensitive to phase, multiplying the frequency amplifies the time error, granting us the ability to achieve proportionally higher precision for frequency distribution. This enhancement in precision holds great promise for various applications that require accurate synchronization over long distances.

\begin{figure}[H]
	\centering
	\includegraphics[width=0.4\textwidth]{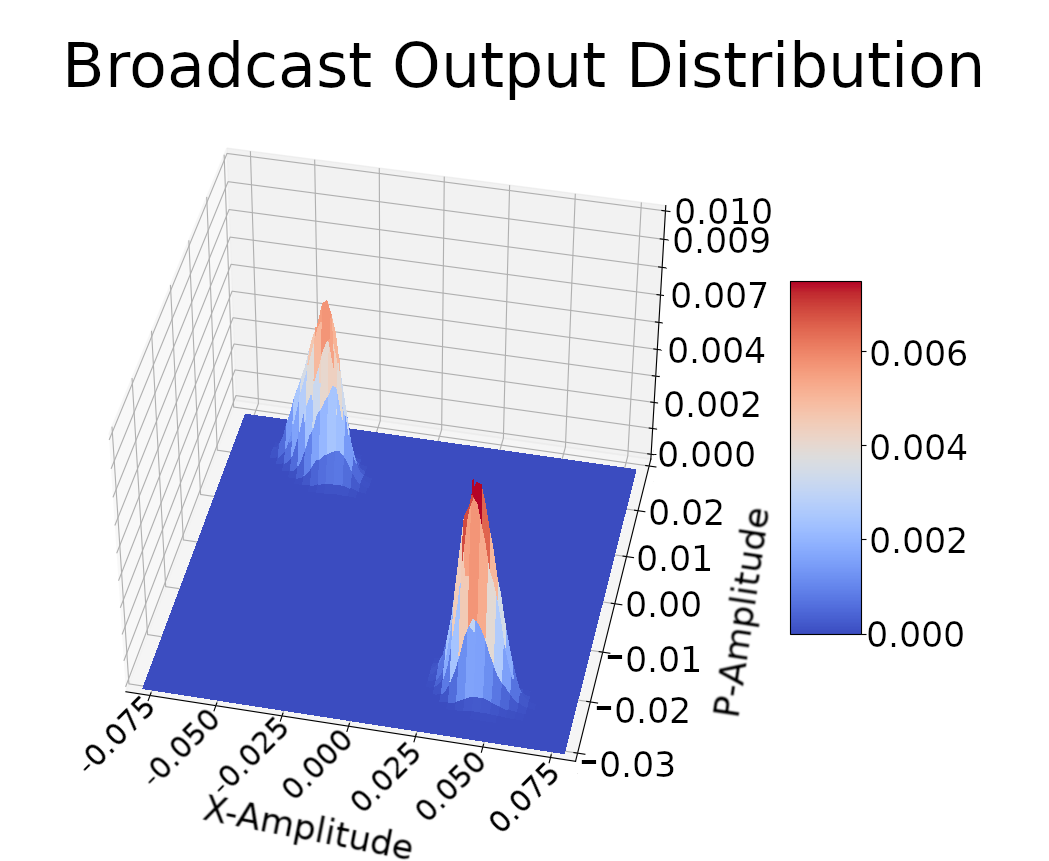}
	\caption{A sample measurement distribution for a broadcast provided by GNURadio \cite{GNURadio}. A phase difference between the measurement cluster centers and the X-axis due to the rotations that this experiment is based on is clearly visible.} \label{fig:BPSK}
\end{figure}

The broadcasts to be performed consist of a pair of thermal distributions. For a plot in phase space, characterised by:

\begin{equation}
 \Psi\left(t\right) = X\cos\left(2\pi f t\right) + Pi\sin\left(2\pi ft\right),
\end{equation}

the measurement distribution is a pair of clusters centered on two points:

\begin{equation}
 \Psi\left(t\right) = A\cos\left(2\pi ft + n\pi\right)~~~~~~ n\in\{0,1\},
\end{equation}

as shown in Figure \ref{fig:BPSK}. A phase error is visible in this Figure, which shows the clusters rotated from an ideal position on the X-axis, caused by differences of frequencies between clocks used as references for the broadcasting and receiving parties. We omit typical forms of phase error correction used in broadcasting, as these are methods of signal processing which have no effect on the reference clocks. Using a feedback controller instead to adjust the frequency of a secondary clock in response to phase errors allows for clock synchronisation.

Taking repeat measurements over the duration of a broadcast between a transmitter and receiver, using a primary and secondary clock respectively, we are able to reconstruct the state as frequency errors cause the measurement clusters to rotate about a centre point between them. For a difference in frequency $\Delta f$ between the two reference clocks, we see a rate of change of phase $\frac{\Delta\phi}{\Delta t}$ given by:

\begin{equation}
 \frac{\Delta\phi}{\Delta t} = 2\pi\Delta f. \label{drift}
\end{equation}

For an ideal, perfectly synchronised pair of clocks, we expect no change in phase over time, therefore we are limited by the minimum detectable phase difference. Using the atomic clocks as references for a higher frequency broadcast allows measurement of changes in broadcast frequencies instead, which magnifies the phase errors, reducing the minimum size of phase errors that can be observed. For some broadcast with a frequency scaled up to $f_{Broadcast}=a\times f_{clock}$, where $f_{clock}=10$ MHz, we instead see a rate of phase drift of:

\begin{equation}
 \frac{\Delta\phi_a}{\Delta t} = 2\pi\Delta f_{Broadcast}=2\pi a~\Delta f_{clock}. \label{drift2}
\end{equation}

Given that radio equipment can be used to broadcast signals far in excess of 10 MHz, this provides an easily adjustable variable which can be increased to improve the minimum detectable frequency errors between a pair of clocks. The limiting factor of this method then becomes the inherent stability of the clocks, and the responsiveness of the feedback controller used to correct the phase errors.

\subsection*{The Wired Channel}

\begin{figure}[H]
	\centering
        \includegraphics[width=0.55\textwidth]{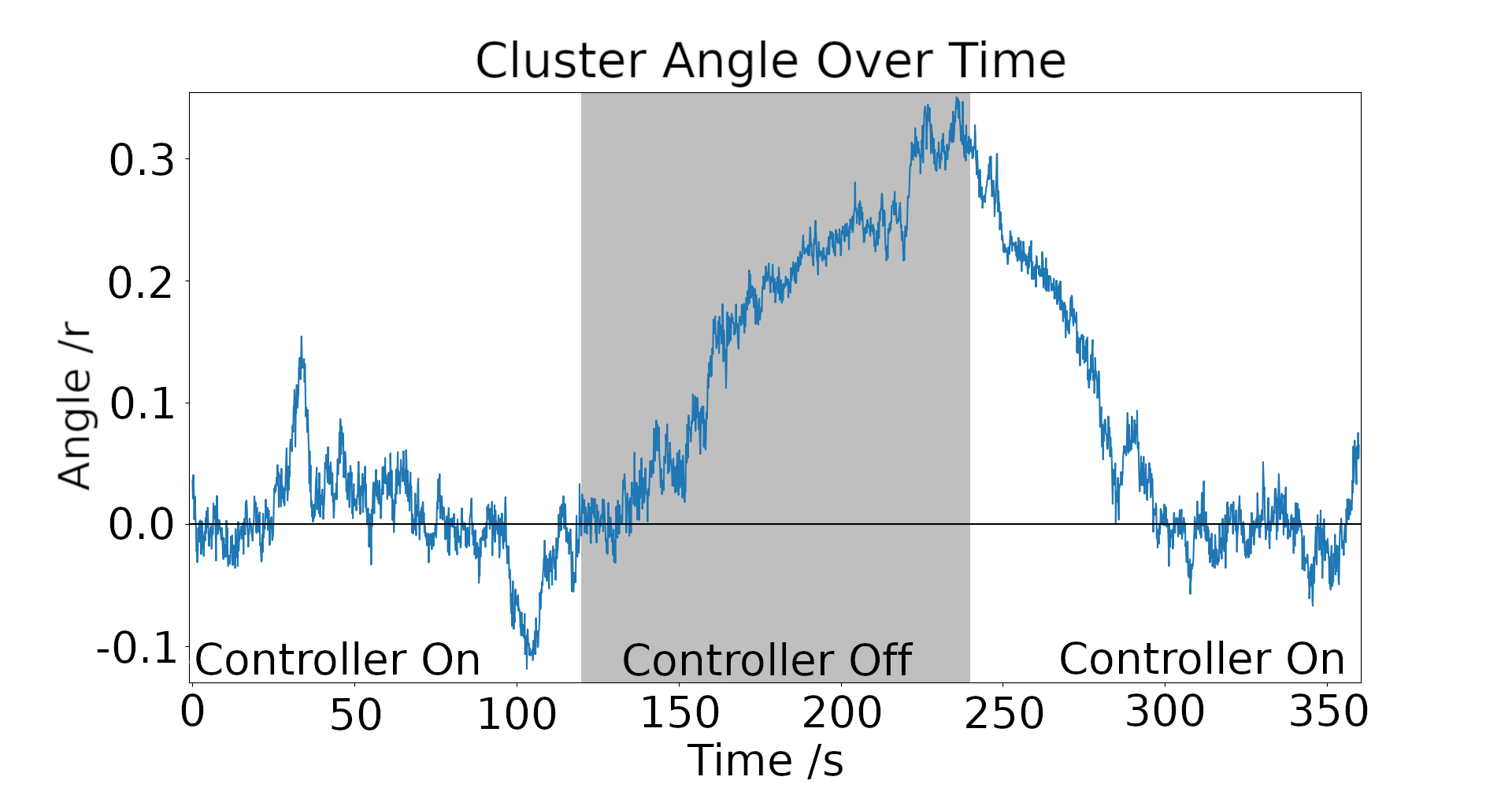}
	\caption{Tracking the average phase of one cluster in a 70 MHz broadcast. The feedback controller was disabled for a period to demonstrate it's effects.} \label{fig:70MHz}
\end{figure}

A test with a wired 70 MHz broadcast is shown in Figure \ref{fig:70MHz}. The Figure shows the effect of the feedback controller on the phase of the reconstructed state, in which the broadcast is performed with the feedback controller for 120 seconds before the controller is disabled for an additional 120 seconds. During this period, phase errors accumulated due to lingering frequency differences. Enabling the controller again readjusts the secondary clock frequency, rotating the measurement cluster back towards a mean of zero phase, where it is maintained. At this point the clocks are once again stabilised.\\

\subsection*{The Wireless Channel}

\begin{figure}[H]
	\centering
	\includegraphics[width=0.45\textwidth]{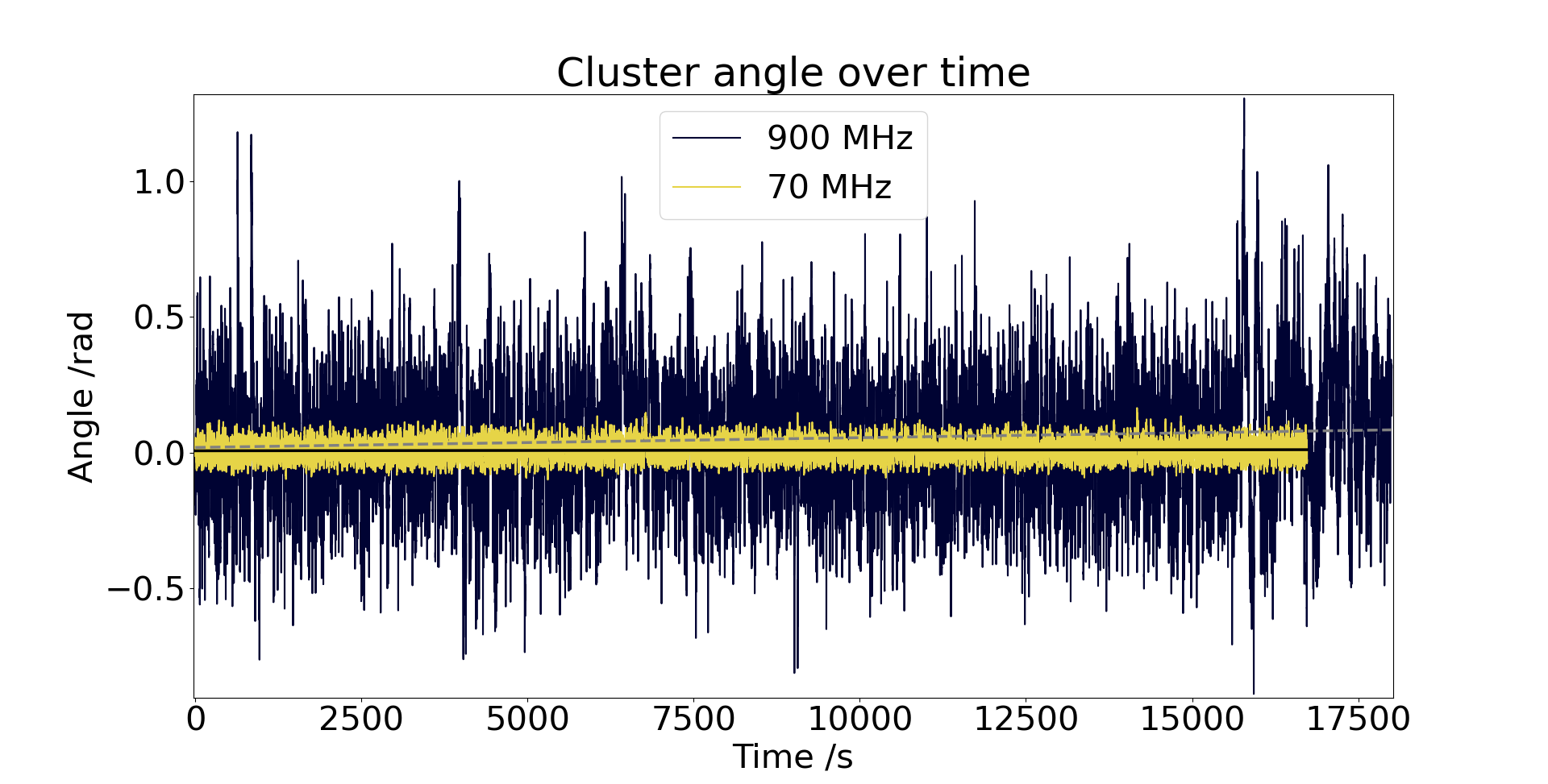}
        \includegraphics[width=0.45\textwidth]{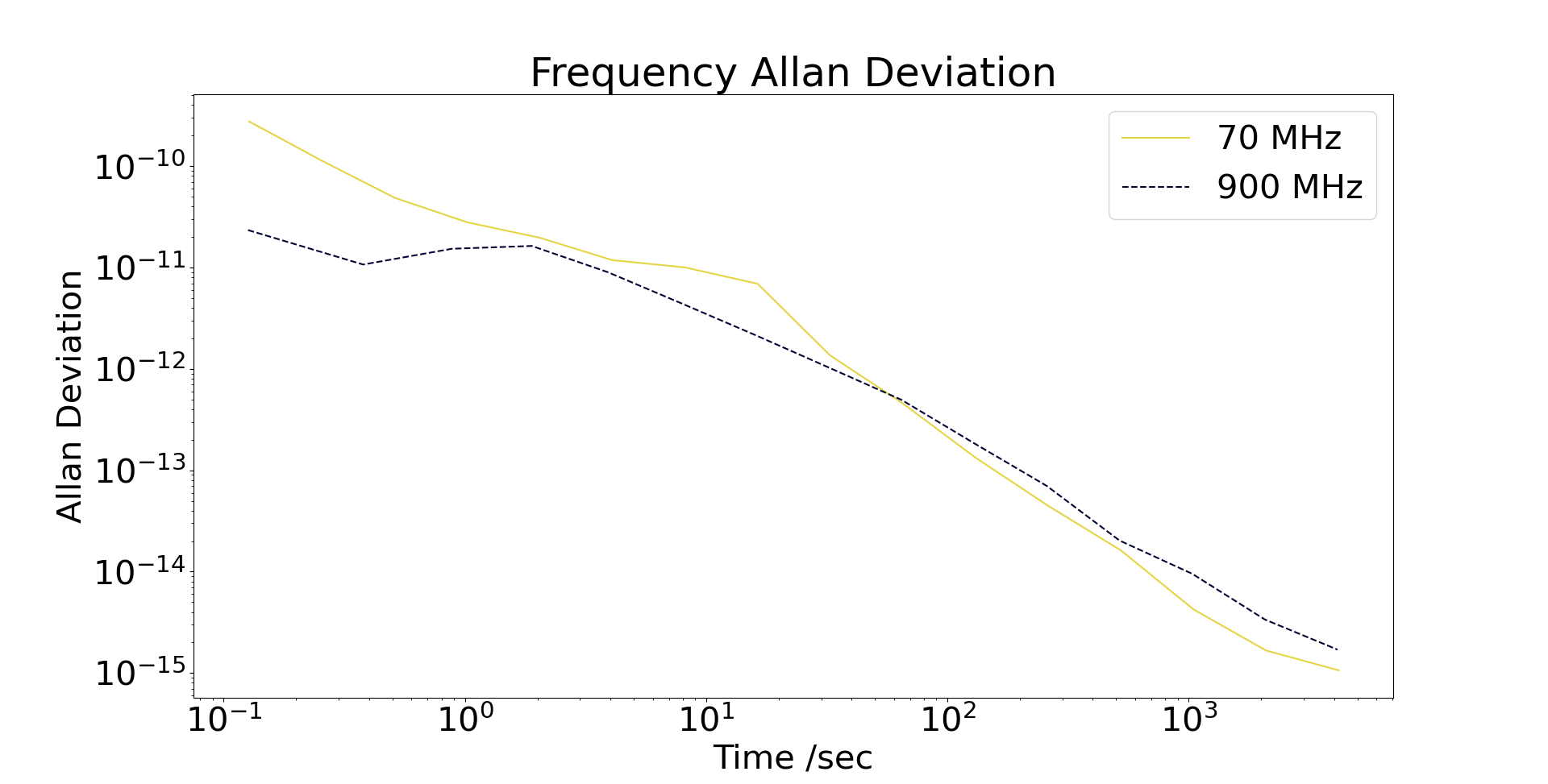}
	\caption{Tracking the average phase of one cluster in a 900 MHz wireless broadcast, and a 70 MHz wired broadcast. Due to the higher frequency, greater instability in the phase drift is visible in the wireless broadcast than wired. The Allan variance for both broadcasts are also shown.} \label{fig:900MHz}
\end{figure}

We then replace the wired channel between the pair of radios with a 2.5 meter wireless channel, and increase the broadcast frequency to 900 MHz. This is an increase in frequency of a factor of 12.9 compared to the 70 MHz wired test, and the drift in phase caused by differences in the reference frequencies increases by the same factor and the gain of the feedback loop therefore also increases. 
The results of this are shown in Figure \ref{fig:900MHz}, accompanied by an Allan deviation plot derived from the same data. These results are shown alongside the measurements from a wired broadcast at 70 MHz for comparison. In this figure you can see the increased phase noise arising from the increase in frequency, however for a longer averaging duration the Allan variance was not noticeably affected by the change from wired to short range wireless broadcasting. In both cases, a small peak is seen in the Allan variance plot, this is likely displaying the delay before feedback from the controller affects the broadcast.
A linear fit shows that the total average rate of change of phase for the wireless broadcast is estimated to be $3.609\times10^{-06} \pm 1.192\times10^{-7}$, which corresponds to a frequency error of 574.3~nHz. Given the source frequency of 900 MHz this corresponds to a free space fractional frequency drift $\frac{\Delta f}{f}=6.382\times 10^{-16}$ accumulated over a 5 hour scan time. 

Using this wireless data we can also plot the modified Allan variance showing $1.701\times 10^{-15}$ at 4136 seconds. These measurements are an improvement on the Allan variance provided in the datasheet for the PRS10 clock used in this test for most averaging times, with the Allan variance never exceeding a part in $10^{13}$ even at the clock's ideal averaging period \cite{PRS10C}. A longer broadcasting time would be required in order to locate the point at which the Allan variance no longer improves, though this will also be dependent on clock performance and the chosen frequency. While the wired broadcast did show an improved Allan deviation compared to the wireless broadcast for a longer averaging duration, demonstrating wireless synchronisation is of greater relevance.\\

Synchronisation was also achieved at 2.4 GHz, a frequency commonly used in communications equipment, though the increased sensitivity to errors at higher frequencies meant that the instability of the clocks used caused less reliable broadcasts. Clocks with improved short-term stability would make higher frequency testing more reliable.

This method only relies on radio broadcasts, which are already employed over long distances and at a wide array of frequencies due to their ubiquitous use in communications, and a suitably responsive feedback controller, which are widely used in a variety of applications and are well understood. This provides clear avenues though which the experimental setup employed here may be improved, through use of broadcasting equipment suitable for longer ranges, and through increasing the broadcast frequency to magnify the effects of clock frequency errors. A clear next step would also be to perform tests with more stable clocks, ideally focusing on a superior primary clock. This is because instability in the primary and secondary clocks are not differentiated between with this method, and are both ascribed to errors in the secondary clock. This may result in errors in stability estimates.\\

Fractional error estimates for frequency measurements in the order of $10^{-16}$ were observed, an improvement of approximately a factor of $10^{3}$ over the minimum advertised Allan variance for the clocks used. This was performed across 2.5m of free space using easily accessible radio equipment, and offers obvious avenues for improvement through longer ranges, higher frequency broadcasts, as well as replacement of the PRS10C clocks with time references that have superior short-term stability.

\subsection*{Author Information}
Adam Walton and Benjamin T. H. Varcoe: These authors contributed equally to the work.

\bibliography{apssamp}

\begin{thebibliography}{19}%
\makeatletter
\providecommand \@ifxundefined [1]{%
 \@ifx{#1\undefined}
}%
\providecommand \@ifnum [1]{%
 \ifnum #1\expandafter \@firstoftwo
 \else \expandafter \@secondoftwo
 \fi
}%
\providecommand \@ifx [1]{%
 \ifx #1\expandafter \@firstoftwo
 \else \expandafter \@secondoftwo
 \fi
}%
\providecommand \natexlab [1]{#1}%
\providecommand \enquote  [1]{``#1''}%
\providecommand \bibnamefont  [1]{#1}%
\providecommand \bibfnamefont [1]{#1}%
\providecommand \citenamefont [1]{#1}%
\providecommand \href@noop [0]{\@secondoftwo}%
\providecommand \href [0]{\begingroup \@sanitize@url \@href}%
\providecommand \@href[1]{\@@startlink{#1}\@@href}%
\providecommand \@@href[1]{\endgroup#1\@@endlink}%
\providecommand \@sanitize@url [0]{\catcode `\\12\catcode `\$12\catcode
  `\&12\catcode `\#12\catcode `\^12\catcode `\_12\catcode `\%12\relax}%
\providecommand \@@startlink[1]{}%
\providecommand \@@endlink[0]{}%
\providecommand \url  [0]{\begingroup\@sanitize@url \@url }%
\providecommand \@url [1]{\endgroup\@href {#1}{\urlprefix }}%
\providecommand \urlprefix  [0]{URL }%
\providecommand \Eprint [0]{\href }%
\providecommand \doibase [0]{https://doi.org/}%
\providecommand \selectlanguage [0]{\@gobble}%
\providecommand \bibinfo  [0]{\@secondoftwo}%
\providecommand \bibfield  [0]{\@secondoftwo}%
\providecommand \translation [1]{[#1]}%
\providecommand \BibitemOpen [0]{}%
\providecommand \bibitemStop [0]{}%
\providecommand \bibitemNoStop [0]{.\EOS\space}%
\providecommand \EOS [0]{\spacefactor3000\relax}%
\providecommand \BibitemShut  [1]{\csname bibitem#1\endcsname}%
\let\auto@bib@innerbib\@empty
\bibitem [{\citenamefont {Tian}\ \emph {et~al.}(2017)\citenamefont {Tian},
  \citenamefont {Song}, \citenamefont {Yu}, \citenamefont {Shi},\ and\
  \citenamefont {Hu}}]{1020Clock}%
  \BibitemOpen
  \bibfield  {author} {\bibinfo {author} {\bibfnamefont {H.}~\bibnamefont
  {Tian}}, \bibinfo {author} {\bibfnamefont {Y.}~\bibnamefont {Song}}, \bibinfo
  {author} {\bibfnamefont {J.}~\bibnamefont {Yu}}, \bibinfo {author}
  {\bibfnamefont {H.}~\bibnamefont {Shi}},\ and\ \bibinfo {author}
  {\bibfnamefont {M.}~\bibnamefont {Hu}},\ }\href
  {https://doi.org/10.1109/JPHOT.2017.2756909} {\bibfield  {journal} {\bibinfo
  {journal} {IEEE Photonics Journal}\ }\textbf {\bibinfo {volume} {9}},\
  \bibinfo {pages} {1} (\bibinfo {year} {2017})}\BibitemShut {NoStop}%
\bibitem [{GPS(2022)}]{GPS}%
  \BibitemOpen
  \href@noop {} {\bibinfo {title} {Gps accuracy}},\ \bibinfo {howpublished}
  {\url{https://www.gps.gov/systems/gps/performance/accuracy/}} (\bibinfo
  {year} {2022}),\ \bibinfo {note} {accessed: 15/12/23}\BibitemShut {NoStop}%
\bibitem [{\citenamefont {Gozzard}(2023)}]{Limit}%
  \BibitemOpen
  \bibfield  {author} {\bibinfo {author} {\bibfnamefont {D.}~\bibnamefont
  {Gozzard}},\ }\href {https://doi.org/10.1038/d41586-023-01937-7} {\bibfield
  {journal} {\bibinfo  {journal} {Nature}\ }\textbf {\bibinfo {volume} {618}},\
  \bibinfo {pages} {680} (\bibinfo {year} {2023})}\BibitemShut {NoStop}%
\bibitem [{\citenamefont {Ludlow}\ \emph {et~al.}(2015)\citenamefont {Ludlow},
  \citenamefont {Boyd}, \citenamefont {Ye}, \citenamefont {Peik},\ and\
  \citenamefont {Schmidt}}]{Review}%
  \BibitemOpen
  \bibfield  {author} {\bibinfo {author} {\bibfnamefont {A.~D.}\ \bibnamefont
  {Ludlow}}, \bibinfo {author} {\bibfnamefont {M.~M.}\ \bibnamefont {Boyd}},
  \bibinfo {author} {\bibfnamefont {J.}~\bibnamefont {Ye}}, \bibinfo {author}
  {\bibfnamefont {E.}~\bibnamefont {Peik}},\ and\ \bibinfo {author}
  {\bibfnamefont {P.~O.}\ \bibnamefont {Schmidt}},\ }\href
  {https://doi.org/10.1103/RevModPhys.87.637} {\bibfield  {journal} {\bibinfo
  {journal} {Rev. Mod. Phys.}\ }\textbf {\bibinfo {volume} {87}},\ \bibinfo
  {pages} {637} (\bibinfo {year} {2015})}\BibitemShut {NoStop}%
\bibitem [{\citenamefont {McGrew}\ \emph {et~al.}(2018)\citenamefont {McGrew},
  \citenamefont {Zhang}, \citenamefont {Fasano}, \citenamefont
  {Sch{\"{a}}ffer}, \citenamefont {Beloy}, \citenamefont {Nicolodi},
  \citenamefont {Brown}, \citenamefont {Hinkley}, \citenamefont {Milani},
  \citenamefont {Schioppo}, \citenamefont {Yoon},\ and\ \citenamefont
  {Ludlow}}]{McGrew2018}%
  \BibitemOpen
  \bibfield  {author} {\bibinfo {author} {\bibfnamefont {W.~F.}\ \bibnamefont
  {McGrew}}, \bibinfo {author} {\bibfnamefont {X.}~\bibnamefont {Zhang}},
  \bibinfo {author} {\bibfnamefont {R.~J.}\ \bibnamefont {Fasano}}, \bibinfo
  {author} {\bibfnamefont {S.~A.}\ \bibnamefont {Sch{\"{a}}ffer}}, \bibinfo
  {author} {\bibfnamefont {K.}~\bibnamefont {Beloy}}, \bibinfo {author}
  {\bibfnamefont {D.}~\bibnamefont {Nicolodi}}, \bibinfo {author}
  {\bibfnamefont {R.~C.}\ \bibnamefont {Brown}}, \bibinfo {author}
  {\bibfnamefont {N.}~\bibnamefont {Hinkley}}, \bibinfo {author} {\bibfnamefont
  {G.}~\bibnamefont {Milani}}, \bibinfo {author} {\bibfnamefont
  {M.}~\bibnamefont {Schioppo}}, \bibinfo {author} {\bibfnamefont {T.~H.}\
  \bibnamefont {Yoon}},\ and\ \bibinfo {author} {\bibfnamefont {A.~D.}\
  \bibnamefont {Ludlow}},\ }\href {https://doi.org/10.1038/s41586-018-0738-2}
  {\bibfield  {journal} {\bibinfo  {journal} {Nature}\ }\textbf {\bibinfo
  {volume} {564}},\ \bibinfo {pages} {87} (\bibinfo {year} {2018})}\BibitemShut
  {NoStop}%
\bibitem [{\citenamefont {Mehlst{\"{a}}ubler}\ \emph
  {et~al.}(2018)\citenamefont {Mehlst{\"{a}}ubler}, \citenamefont {Grosche},
  \citenamefont {Lisdat}, \citenamefont {Schmidt},\ and\ \citenamefont
  {Denker}}]{Mehlstaubler2018}%
  \BibitemOpen
  \bibfield  {author} {\bibinfo {author} {\bibfnamefont {T.~E.}\ \bibnamefont
  {Mehlst{\"{a}}ubler}}, \bibinfo {author} {\bibfnamefont {G.}~\bibnamefont
  {Grosche}}, \bibinfo {author} {\bibfnamefont {C.}~\bibnamefont {Lisdat}},
  \bibinfo {author} {\bibfnamefont {P.~O.}\ \bibnamefont {Schmidt}},\ and\
  \bibinfo {author} {\bibfnamefont {H.}~\bibnamefont {Denker}},\ }\href
  {https://doi.org/10.1088/1361-6633/aab409} {\bibfield  {journal} {\bibinfo
  {journal} {Reports on Progress in Physics}\ }\textbf {\bibinfo {volume}
  {81}},\ \bibinfo {pages} {64401} (\bibinfo {year} {2018})}\BibitemShut
  {NoStop}%
\bibitem [{\citenamefont {Riehle}(2017)}]{Riehle2017}%
  \BibitemOpen
  \bibfield  {author} {\bibinfo {author} {\bibfnamefont {F.}~\bibnamefont
  {Riehle}},\ }\href {https://doi.org/10.1038/nphoton.2016.235} {\bibfield
  {journal} {\bibinfo  {journal} {Nature Photonics}\ }\textbf {\bibinfo
  {volume} {11}},\ \bibinfo {pages} {25} (\bibinfo {year} {2017})}\BibitemShut
  {NoStop}%
\bibitem [{\citenamefont {G.~Barontini}(2022)}]{QSNET}%
  \BibitemOpen
  \bibfield  {author} {\bibinfo {author} {\bibfnamefont {V.~B.}\ \bibnamefont
  {G.~Barontini}, \bibfnamefont {L.~Blackburn}},\ }\bibfield  {journal}
  {\bibinfo  {journal} {EPJ Quantum Technology}\ }\textbf {\bibinfo {volume}
  {9}},\ \href {https://doi.org/10.1140/epjqt/s40507-022-00130-5}
  {10.1140/epjqt/s40507-022-00130-5} (\bibinfo {year} {2022})\BibitemShut
  {NoStop}%
\bibitem [{\citenamefont {Kolkowitz}\ \emph {et~al.}(2016)\citenamefont
  {Kolkowitz}, \citenamefont {Pikovski}, \citenamefont {Langellier},
  \citenamefont {Lukin}, \citenamefont {Walsworth},\ and\ \citenamefont
  {Ye}}]{Waves}%
  \BibitemOpen
  \bibfield  {author} {\bibinfo {author} {\bibfnamefont {S.}~\bibnamefont
  {Kolkowitz}}, \bibinfo {author} {\bibfnamefont {I.}~\bibnamefont {Pikovski}},
  \bibinfo {author} {\bibfnamefont {N.}~\bibnamefont {Langellier}}, \bibinfo
  {author} {\bibfnamefont {M.~D.}\ \bibnamefont {Lukin}}, \bibinfo {author}
  {\bibfnamefont {R.~L.}\ \bibnamefont {Walsworth}},\ and\ \bibinfo {author}
  {\bibfnamefont {J.}~\bibnamefont {Ye}},\ }\href
  {https://doi.org/10.1103/PhysRevD.94.124043} {\bibfield  {journal} {\bibinfo
  {journal} {Phys. Rev. D}\ }\textbf {\bibinfo {volume} {94}},\ \bibinfo
  {pages} {124043} (\bibinfo {year} {2016})}\BibitemShut {NoStop}%
\bibitem [{\citenamefont {Wcis{\l}o}\ \emph {et~al.}(2016)\citenamefont
  {Wcis{\l}o}, \citenamefont {Morzy{\'{n}}ski}, \citenamefont {Bober},
  \citenamefont {Cygan}, \citenamefont {Lisak}, \citenamefont {Ciury{\l}o},\
  and\ \citenamefont {Zawada}}]{Dark}%
  \BibitemOpen
  \bibfield  {author} {\bibinfo {author} {\bibfnamefont {P.}~\bibnamefont
  {Wcis{\l}o}}, \bibinfo {author} {\bibfnamefont {P.}~\bibnamefont
  {Morzy{\'{n}}ski}}, \bibinfo {author} {\bibfnamefont {M.}~\bibnamefont
  {Bober}}, \bibinfo {author} {\bibfnamefont {A.}~\bibnamefont {Cygan}},
  \bibinfo {author} {\bibfnamefont {D.}~\bibnamefont {Lisak}}, \bibinfo
  {author} {\bibfnamefont {R.}~\bibnamefont {Ciury{\l}o}},\ and\ \bibinfo
  {author} {\bibfnamefont {M.}~\bibnamefont {Zawada}},\ }\href
  {https://doi.org/10.1038/s41550-016-0009} {\bibfield  {journal} {\bibinfo
  {journal} {Nature Astronomy}\ }\textbf {\bibinfo {volume} {1}},\ \bibinfo
  {pages} {0009} (\bibinfo {year} {2016})}\BibitemShut {NoStop}%
\bibitem [{NPL(2023)}]{NPL}%
  \BibitemOpen
  \href@noop {} {\bibinfo {title} {Time scales}},\ \bibinfo {howpublished}
  {\url{https://www.npl.co.uk/time-frequency/time-scales}} (\bibinfo {year}
  {2023}),\ \bibinfo {note} {accessed: 04/07/23}\BibitemShut {NoStop}%
\bibitem [{\citenamefont {Cizek}\ \emph {et~al.}(2022)\citenamefont {Cizek},
  \citenamefont {Pravdova}, \citenamefont {Pham}, \citenamefont {Lesundak},
  \citenamefont {Hrabina}, \citenamefont {Lazar}, \citenamefont {Pronebner},
  \citenamefont {Aeikens}, \citenamefont {Premper}, \citenamefont {Havlis},
  \citenamefont {Velc}, \citenamefont {Smotlacha}, \citenamefont {Altmannova},
  \citenamefont {Schumm}, \citenamefont {Vojtech}, \citenamefont {Niessner},\
  and\ \citenamefont {Cip}}]{Cizek:22}%
  \BibitemOpen
  \bibfield  {author} {\bibinfo {author} {\bibfnamefont {M.}~\bibnamefont
  {Cizek}}, \bibinfo {author} {\bibfnamefont {L.}~\bibnamefont {Pravdova}},
  \bibinfo {author} {\bibfnamefont {T.~M.}\ \bibnamefont {Pham}}, \bibinfo
  {author} {\bibfnamefont {A.}~\bibnamefont {Lesundak}}, \bibinfo {author}
  {\bibfnamefont {J.}~\bibnamefont {Hrabina}}, \bibinfo {author} {\bibfnamefont
  {J.}~\bibnamefont {Lazar}}, \bibinfo {author} {\bibfnamefont
  {T.}~\bibnamefont {Pronebner}}, \bibinfo {author} {\bibfnamefont
  {E.}~\bibnamefont {Aeikens}}, \bibinfo {author} {\bibfnamefont
  {J.}~\bibnamefont {Premper}}, \bibinfo {author} {\bibfnamefont
  {O.}~\bibnamefont {Havlis}}, \bibinfo {author} {\bibfnamefont
  {R.}~\bibnamefont {Velc}}, \bibinfo {author} {\bibfnamefont {V.}~\bibnamefont
  {Smotlacha}}, \bibinfo {author} {\bibfnamefont {L.}~\bibnamefont
  {Altmannova}}, \bibinfo {author} {\bibfnamefont {T.}~\bibnamefont {Schumm}},
  \bibinfo {author} {\bibfnamefont {J.}~\bibnamefont {Vojtech}}, \bibinfo
  {author} {\bibfnamefont {A.}~\bibnamefont {Niessner}},\ and\ \bibinfo
  {author} {\bibfnamefont {O.}~\bibnamefont {Cip}},\ }\href
  {https://doi.org/10.1364/OE.447498} {\bibfield  {journal} {\bibinfo
  {journal} {Opt. Express}\ }\textbf {\bibinfo {volume} {30}},\ \bibinfo
  {pages} {5450} (\bibinfo {year} {2022})}\BibitemShut {NoStop}%
\bibitem [{\citenamefont {Caldwell}\ \emph {et~al.}(2023)\citenamefont
  {Caldwell}, \citenamefont {Deschênes}, \citenamefont {Ellis}, \citenamefont
  {Swann}, \citenamefont {Stuhl}, \citenamefont {Bergeron}, \citenamefont
  {Newbury},\ and\ \citenamefont {Sinclair}}]{Geo}%
  \BibitemOpen
  \bibfield  {author} {\bibinfo {author} {\bibfnamefont {E.}~\bibnamefont
  {Caldwell}}, \bibinfo {author} {\bibfnamefont {J.-D.}\ \bibnamefont
  {Deschênes}}, \bibinfo {author} {\bibfnamefont {J.}~\bibnamefont {Ellis}},
  \bibinfo {author} {\bibfnamefont {W.}~\bibnamefont {Swann}}, \bibinfo
  {author} {\bibfnamefont {B.}~\bibnamefont {Stuhl}}, \bibinfo {author}
  {\bibfnamefont {H.}~\bibnamefont {Bergeron}}, \bibinfo {author}
  {\bibfnamefont {N.}~\bibnamefont {Newbury}},\ and\ \bibinfo {author}
  {\bibfnamefont {L.}~\bibnamefont {Sinclair}},\ }\href
  {https://doi.org/10.1038/s41586-023-06032-5} {\bibfield  {journal} {\bibinfo
  {journal} {Nature}\ }\textbf {\bibinfo {volume} {618}},\ \bibinfo {pages}
  {721} (\bibinfo {year} {2023})}\BibitemShut {NoStop}%
\bibitem [{\citenamefont {Desch\^enes}\ \emph {et~al.}(2016)\citenamefont
  {Desch\^enes}, \citenamefont {Sinclair}, \citenamefont {Giorgetta},
  \citenamefont {Swann}, \citenamefont {Baumann}, \citenamefont {Bergeron},
  \citenamefont {Cermak}, \citenamefont {Coddington},\ and\ \citenamefont
  {Newbury}}]{Distant}%
  \BibitemOpen
  \bibfield  {author} {\bibinfo {author} {\bibfnamefont {J.-D.}\ \bibnamefont
  {Desch\^enes}}, \bibinfo {author} {\bibfnamefont {L.~C.}\ \bibnamefont
  {Sinclair}}, \bibinfo {author} {\bibfnamefont {F.~R.}\ \bibnamefont
  {Giorgetta}}, \bibinfo {author} {\bibfnamefont {W.~C.}\ \bibnamefont
  {Swann}}, \bibinfo {author} {\bibfnamefont {E.}~\bibnamefont {Baumann}},
  \bibinfo {author} {\bibfnamefont {H.}~\bibnamefont {Bergeron}}, \bibinfo
  {author} {\bibfnamefont {M.}~\bibnamefont {Cermak}}, \bibinfo {author}
  {\bibfnamefont {I.}~\bibnamefont {Coddington}},\ and\ \bibinfo {author}
  {\bibfnamefont {N.~R.}\ \bibnamefont {Newbury}},\ }\href
  {https://doi.org/10.1103/PhysRevX.6.021016} {\bibfield  {journal} {\bibinfo
  {journal} {Phys. Rev. X}\ }\textbf {\bibinfo {volume} {6}},\ \bibinfo {pages}
  {021016} (\bibinfo {year} {2016})}\BibitemShut {NoStop}%
\bibitem [{\citenamefont {Guo}\ \emph {et~al.}(2022)\citenamefont {Guo},
  \citenamefont {Gao}, \citenamefont {Bai}, \citenamefont {Pan}, \citenamefont
  {Liu}, \citenamefont {Lu},\ and\ \citenamefont {Zhang}}]{Space}%
  \BibitemOpen
  \bibfield  {author} {\bibinfo {author} {\bibfnamefont {Y.}~\bibnamefont
  {Guo}}, \bibinfo {author} {\bibfnamefont {S.}~\bibnamefont {Gao}}, \bibinfo
  {author} {\bibfnamefont {Y.}~\bibnamefont {Bai}}, \bibinfo {author}
  {\bibfnamefont {Z.}~\bibnamefont {Pan}}, \bibinfo {author} {\bibfnamefont
  {Y.}~\bibnamefont {Liu}}, \bibinfo {author} {\bibfnamefont {X.}~\bibnamefont
  {Lu}},\ and\ \bibinfo {author} {\bibfnamefont {S.}~\bibnamefont {Zhang}},\
  }\href {https://doi.org/10.3390/rs14030528} {\bibfield  {journal} {\bibinfo
  {journal} {Remote Sensing}\ }\textbf {\bibinfo {volume} {14}},\ \bibinfo
  {pages} {528} (\bibinfo {year} {2022})}\BibitemShut {NoStop}%
\bibitem [{\citenamefont {Piester}\ \emph {et~al.}(2011)\citenamefont
  {Piester}, \citenamefont {Rost}, \citenamefont {Fujieda}, \citenamefont
  {Feldmann},\ and\ \citenamefont {Bauch}}]{SynchARS}%
  \BibitemOpen
  \bibfield  {author} {\bibinfo {author} {\bibfnamefont {D.}~\bibnamefont
  {Piester}}, \bibinfo {author} {\bibfnamefont {M.}~\bibnamefont {Rost}},
  \bibinfo {author} {\bibfnamefont {M.}~\bibnamefont {Fujieda}}, \bibinfo
  {author} {\bibfnamefont {T.}~\bibnamefont {Feldmann}},\ and\ \bibinfo
  {author} {\bibfnamefont {A.}~\bibnamefont {Bauch}},\ }\href
  {https://doi.org/10.5194/ars-9-1-2011} {\bibfield  {journal} {\bibinfo
  {journal} {Advances in Radio Science}\ }\textbf {\bibinfo {volume} {9}},\
  \bibinfo {pages} {1} (\bibinfo {year} {2011})}\BibitemShut {NoStop}%
\bibitem [{GNS(2023)}]{GNSS}%
  \BibitemOpen
  \href@noop {} {\bibinfo {title} {Gnss and satellite time transfer}},\
  \bibinfo {howpublished}
  {\url{https://www.nist.gov/pml/time-and-frequency-division/time-services/gnss-and-satellite-time-transfer}}
  (\bibinfo {year} {2023}),\ \bibinfo {note} {accessed: 05/07/23}\BibitemShut
  {NoStop}%
\bibitem [{GNU(2023)}]{GNURadio}%
  \BibitemOpen
  \href@noop {} {\bibinfo {title} {Gnuradio}},\ \bibinfo {howpublished}
  {\url{https://www.gnuradio.org}} (\bibinfo {year} {2023}),\ \bibinfo {note}
  {accessed: 09/06/23}\BibitemShut {NoStop}%
\bibitem [{PRS(2023)}]{PRS10C}%
  \BibitemOpen
  \href@noop {} {\bibinfo {title} {Prs10 — rubidium frequency standard with
  low phase noise}},\ \bibinfo {howpublished}
  {\url{https://www.thinksrs.com/downloads/pdfs/catalog/PRS10c.pdf}} (\bibinfo
  {year} {2023}),\ \bibinfo {note} {accessed: 18/07/23}\BibitemShut {NoStop}%
\end{thebibliography}%

\end{document}